\documentclass[prb,twocolumn,showpacs]{revtex4}

\usepackage{graphicx}
\usepackage{subfigure}
\usepackage{amsfonts}
\usepackage{amsmath}
\usepackage{amssymb}

\usepackage{color}

\usepackage{hyperref}

\begin{document}

\title{Static Screening and Delocalization Effects in the Hubbard-Anderson Model}

\author{Peter \surname{Henseler}}
\author{Johann \surname{Kroha}}
\affiliation{Physikalisches Institut, Universit\"at Bonn, Nu\ss allee 12, D-53115 Bonn, Germany}
\author{Boris \surname{Shapiro}}
\affiliation{Department of Physics, Technion-Israel Institute of Technology, Haifa 32000, Israel}

\begin{abstract}
We study the suppression of electron localization due to the screening of
disorder in a Hubbard-Anderson model. We focus on the change of the electron
localization length at the Fermi level within a static picture, where
interactions are absorbed into the redefinition of the random on-site
energies. Two different approximations are presented, either one yielding a
nonmonotonic dependence of the localization length on the interaction
strength, with a pronounced maximum at an intermediate interaction strength. 
In spite of its simplicity, our approach is in good agreement with recent 
numerical results.
\end{abstract}


\maketitle

\section{Introduction}
Understanding the interplay between disorder and electron-electron interactions remains one of the major challenges in modern condensed matter physics, experimental as well as theoretical. The research in this field has been stimulated by the possible metallic behavior in two-dimensional disordered interacting systems.\cite{aks01} A metallic phase at zero temperature in two or less dimensions would be in contrast to the prediction of the scaling theory of Anderson localization.\cite{aalr79} The possible existence of a metallic phase, induced by interactions, is a long standing\cite{fa80} and still controversial problem, discussed by many authors during the last three decades.\cite{fussnote1} \\
One of the ideas proposed and discussed by several authors is that interactions lead to a partial screening of the random potential and, thus, reduce the effect of localization. In particular, the 2$d$ disordered Hubbard model (the Hubbard-Anderson model) has been studied, mostly numerically, and it was demonstrated that repulsive interactions can have a delocalizing effect.\cite{dst99,her01,sbs03,ht04,cds07,swa07,np05} \\
In this paper, we present an analytical study of the screening effect, focussing on the case of strong disorder at zero temperature, when the Hubbard-Anderson model is in the regime of an Anderson insulator. Our approach is based on an exact treatment of the atomic limit, followed by "switching on" the intersite hopping $t$, under the assumption that the atomic-limit occupation numbers do not change. Let us emphasize that in our static approach, the interactions only change the original on-site energies. Therefore, we are left with a single-particle Anderson Hamiltonian, and the question is how the localized single-particle states may change due to the new, renormalized probability distribution of the on-site energies. In this sense, the approach is close, although not identical, to the Hartree-Fock treatment. A comparison of both methods will be presented. Although this approach is formulated for an arbitrary filling factor, it becomes inadequate close to half filling where magnetic effects dominate\cite{adak07} (such effects are not considered in our work).\\
Furthermore, we will show that, for fixed disorder, the localization length $\xi$ is a nonmonotonic function of the Hubbard interaction energy $U$, with a maximum for some intermediate value of $U$. 
This is because for strong interactions, the Mott-Hubbard physics of interaction-suppressed hopping dominates, leading to the formation of two disorder broadened Hubbard bands with a reduced average density of states at the Fermi level and, as a consequence, increasing effectively the disorder strength.
In spite of the simplicity of the approach, the results are in good agreement with recent numerical studies,\cite{sbs03,cds07,tdhs05,ds03,swa07,dst99,ht04} in the appropriate range of parameters. \\
The evaluation of the localization length within our present study is limited to one dimension. However, since the competition of screening on the one hand and Mott-Hubbard physics on the other hand operates in any dimension, the nonmonotonic dependence of $\xi$ on $U$ should also hold in two and three dimensions, as argued below.

\section{Atomic-Limit Approximation}
In this paper, we consider the Hubbard-Anderson model, with on-site repulsion and on-site disorder, at zero temperature. The corresponding Hamiltonian is 
\begin{eqnarray}\label{original Hamiltonian}
 H & = & H_{0} \: + \: H_{\textrm{kin}} \: + \: H_{\textrm{e-e}} \nonumber \\ 
 & = & \sum\limits_{i,\sigma} \left(\varepsilon_{i}^{\phantom{\dagger}}-\mu\right) c_{i\sigma}^{\dagger} c_{i\sigma}^{\phantom{\dagger}} \: - \: t\!\sum\limits_{<i,j>,\sigma} \!c_{i,\sigma}^{\dagger} c_{j\sigma}^{\phantom{\dagger}} \nonumber \\
 & & \: + \: U \sum\limits_{i} n_{i\uparrow}^{\phantom{\dagger}} n_{i\downarrow}^{\phantom{\dagger}} \, .	
\end{eqnarray}
\begin{figure}
 \centering
 \includegraphics[scale=0.35]{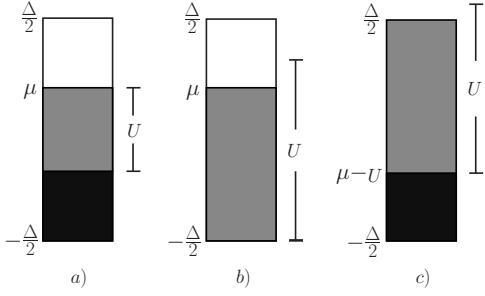}
 \caption{Site occupation in the atomic ground state with doubly occupied (black), singly occupied (gray) and empty states (white). (a) weak interaction, (b) strong interaction for less than half filling (i.e., $\rho<1$), and (c) strong interaction for more than half filling $(\rho>1)$.}\label{fig chem pot}
\end{figure}\noindent
As usual, $c_{i\sigma}^{\dagger} (c_{i\sigma}^{\phantom{\dagger}})$ denote
fermion creation (destruction) operators of an electron at site $i$ with spin
$\sigma$, $n_{i\sigma}^{\phantom{\dagger}}=c_{i\sigma}^{\dagger}
c_{i\sigma}^{\phantom{\dagger}}$, $t$ is the nearest-neighbor hopping
amplitude, $U$ is the on-site repulsion, $\mu$ is the chemical potential, and
$\{\varepsilon_{i}^{\phantom{\dagger}}\}$ are the on-site energies. The latter
are assumed to be independent and uniformly distributed over the interval
$[-\frac{\Delta}{2},\frac{\Delta}{2}]$,
with the disorder parameter $\Delta$. To focus on the screening effect in the
case of strong localization, i.e., $\Delta\gg t$, the interaction term will be
absorbed into the on-site energies, yielding a renormalized distribution of
the $\varepsilon_{i}$. This results in an effective single-particle problem
with a probability function $p_{\textrm{A}}(\varepsilon_{i})$ which is
derived as follows. \\ 
In the atomic limit ($t=0$), the ground state of the system can be solved
exactly for an arbitrary filling factor
$\rho=\frac{\textrm{N}_{\textrm{e}}}{\textrm{N}}$, where
$\textrm{N}_{\textrm{e}}$, $\textrm{N}$ are the numbers of electrons and
  lattice sites, respectively.\cite{cds07} The chemical potential $\mu$ and
  the site-dependent occupation numbers $\langle n_{i}\rangle_{0}$ can be
  expressed as functions of $\rho$, $\Delta$, and $U$: All sites with on-site
  energies below $\mu-U$ are doubly occupied, all sites within $[\mu-U,\mu]$
  are singly occupied, and all other sites are empty (see Fig. \ref{fig chem
    pot}). Thus, the total occupation number of site $i$ and the 
chemical potential are, respectively,  
\begin{eqnarray}\label{eqn occupation numbers atomic limit}
\langle n_{i} \rangle_{0} & = & 
 \left\{ \begin{array}{rcl} 2 & , &
  \varepsilon_{i} \leq \mu - U \\ 1 & , & 
 \mu - U < \varepsilon_{i}
  \leq \mu \\ 0 & , & \varepsilon_{i} > \mu \end{array} \right. 
\end{eqnarray}
and\cite{fussnote4}
\begin{eqnarray}
\label{eqn chem pot}
 \mu & = & \left\{ \begin{array}{rcl} \frac{1}{2}(\Delta\rho - \Delta + U) & , & \rho < 1 \, , \, U < \Delta\rho  \\ \Delta (\rho - \frac{1}{2}) & , & \rho < 1 \, , \, U \geq \Delta\rho  \\ \frac{1}{2}(\Delta\rho - \Delta + U) & , & \rho \geq 1 \, , \, U < 2\Delta - \Delta\rho \\ \Delta(\rho-\frac{3}{2})+U & , & \rho \geq 1 \, , \, U \geq 2\Delta - \Delta\rho. \end{array} \right. 
\end{eqnarray}
The renormalized site energies depend on the site occupation numbers and can
be read off from the poles of the (time-ordered) single-particle propagator,
\begin{eqnarray}\label{eqn propagator}
 & & G_{i\sigma}(\omega) \: = \: \frac{ \langle n_{i,\sigma} \rangle_{0} \langle n_{i,-\sigma} \rangle_{0}}{\omega-(\varepsilon_{i}-\mu +U) -i0^{+}} \nonumber \\ 
 & & \: + \: \frac{ \langle n_{i,\sigma} \rangle_{0} (1- \langle n_{i,-\sigma} \rangle_{0} )}{\omega-(\varepsilon_{i}-\mu) -i0^{+}} \: + \: \frac{ (1- \langle n_{i,\sigma} \rangle_{0} ) \langle n_{i,-\sigma} \rangle_{0}}{\omega-(\varepsilon_{i}-\mu +U) +i0^{+}}  \nonumber\\ 
& & \: + \: \frac{ (1- \langle n_{i,\sigma} \rangle_{0}) (1- \langle n_{i,-\sigma} \rangle_{0} )}{\omega-(\varepsilon_{i}-\mu) +i0^{+}} \quad .
\end{eqnarray}
The first term in Eq.~(\ref{eqn propagator}) corresponds to the doubly occupied
sites showing that these on-site energies are shifted by  $U$. The next two
terms correspond to singly occupied sites. In the absence of spin
polarization, $\langle n_{i,\sigma} \rangle_{0} = \langle n_{i,-\sigma}
\rangle_{0}$, half of these on-site energies are again shifted by $U$ whereas
the other half remain unchanged. Finally, the last term corresponds to the
unoccupied sites, whose energies also remain unchanged. 
Combining Eqs.~(\ref{eqn chem pot}) and (\ref{eqn propagator}), the rule for
replacing the bare site energy $\varepsilon_i$ by a renormalized one is
\begin{eqnarray}
 \varepsilon_{i}^{\phantom{\dagger}} & \mapsto & \!\!\left\{\!\!
 \begin{array}{rcl} \varepsilon_{i}^{\phantom{\dagger}}+U \, &
 \mbox{if} & \varepsilon_{i}^{\phantom{\dagger}} \leq \mu - U \\
 \begin{array}{r} \varepsilon_{i}^{\phantom{\dagger}} + U \\
 \varepsilon_{i}^{\phantom{\dagger}} \end{array} & \mbox{if} & \mu -
 U < \varepsilon_{i}^{\phantom{\dagger}} \leq \mu \: \left( \begin{array}{l} \mbox{each with} \\ \mbox{prob. of $\frac{1}{2}$} \end{array} \right) \\ \varepsilon_{i}^{\phantom{\dagger}} \, &  \mbox{if} & \varepsilon_{i}^{\phantom{\dagger}} > \mu. \end{array} \right. \quad
\end{eqnarray}
\begin{figure}
 \begin{center}
  \subfigure[]{\includegraphics[scale=0.53]{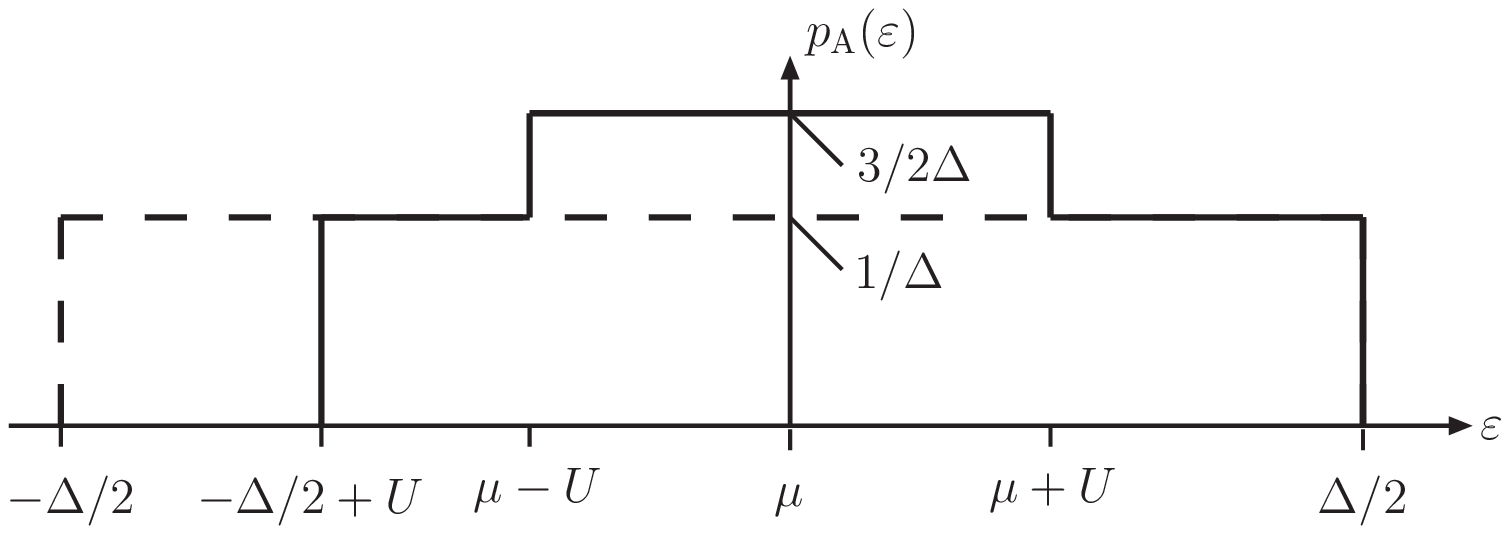} \label{fig prob distribution atomic}}
  \hfil
  \subfigure[]{\includegraphics[scale=0.53]{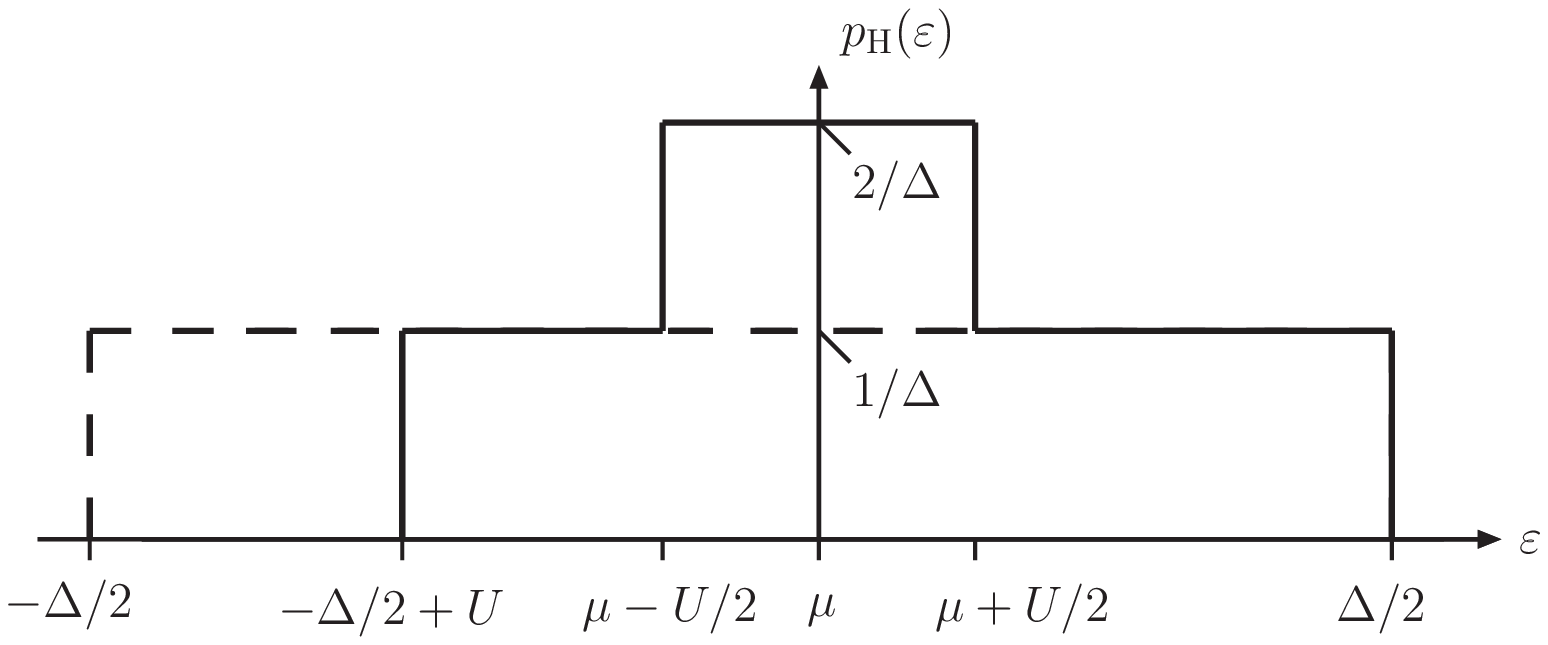} \label{fig prob distribution hartree}}
  \caption{The renormalized on-site energy probability functions (solid line) for weak repulsion $U$: (a) atomic-limit approximation, and (b) Hartree-Fock approximation. For comparison, also the original function is shown (dashed lines).}\label{fig prob distribution}
 \end{center}
\end{figure} 
\noindent For a weak to intermediate repulsion $U$ these shifts lead to a raise of the lowest lying on-site energies towards the Fermi level $\mu$ resulting in a renormalized probability function $p_{\textrm{A}}(\varepsilon)$ with a reduced width and modified shape (see Fig. \ref{fig prob distribution atomic}). \\
In the atomic-limit approximation the Hamiltonian Eq.~(\ref{original Hamiltonian}) is replaced by the effective single-particle Anderson Hamiltonian 
\begin{eqnarray}\label{atomic Hamiltonian}
 H & = & \sum\limits_{i\sigma} \!\left(\varepsilon_{i}^{\phantom{\dagger}}-\mu\right) c_{i\sigma}^{\dagger} c_{i\sigma}^{\phantom{\dagger}}  -  t\sum\limits_{<i,j>\sigma} c_{i\sigma}^{\dagger} c_{j\sigma}^{\phantom{\dagger}} 
\end{eqnarray}
with on-site energy probability function $p_{\textrm{A}}(\varepsilon)$, where the two-particle interaction $U$ enters only as a (screening) parameter in $p_{\textrm{A}}(\varepsilon)$. Note that, due to the asymmetry of $p_{\textrm{A}}(\varepsilon)$, the average value of the renormalized on-site energies is 
\begin{eqnarray}
 \langle \varepsilon \rangle_{\textrm{A}} & = & \int p_{\textrm{A}}(\varepsilon)\, \varepsilon \, d\varepsilon \: = \: \frac{1}{2}\,\rho\, U \, .
\end{eqnarray}
In deriving Eq.~(\ref{atomic Hamiltonian}), it was assumed that in the case of strong disorder, the occupation numbers in the atomic ground state, Eq.~(\ref{eqn occupation numbers atomic limit}), are close to the occupation numbers $\langle n_{i}\rangle$ in the true ground state, with finite hopping amplitude $t$. This assumption is based on the results from the single-particle theory of localization,\cite{and58} where it is known that for strong disorder, an electron, once located at any site $i$, will stay at that site with high probability. More precisely, the change of the occupation number of site $i$, $\delta\langle n_{i}\rangle \equiv \langle n_{i} \rangle - \langle n_{i} \rangle_{0}$, is of order $t^{2}/\Delta$. The same estimate holds if one considers hopping of a single electron on the background of the other electrons, which are assumed to be immobile. Furthermore, for strong disorder, a perturbative expansion in $t$, around the atomic ground state of Eq.~(\ref{original Hamiltonian}), is possible.\cite{fwgf89} Thus, with the same reasoning as in the noninteracting case, the leading self-energy corrections are again only of second order.  \\
One measure of localization is the localization length $\xi$ which governs the
exponential decay of the single-particle wave functions $\psi(r)$ at distances
far away from its localization center. It can be calculated from the
probability for a transition from site $i$ to site $j$ as\cite{lgp88}
\begin{eqnarray}\label{eqn def loc length}
 -\frac{1}{\xi(E)} & = & \lim_{|x_{i}-x_{j}|\rightarrow \infty} \frac{\log \left\langle \left| G^{\textrm{R}}_{ij}(E) \right|^{2} \right\rangle}{2 |x_{i}-x_{j}|} ,
\end{eqnarray}
where $G^{\textrm{R}}_{ij}(E)$ is the retarded propagator for a particle 
with energy $E$ from site $j$ to site $i$.
In general, it is not possible to deduce $\xi$ from the probability
distribution analytically, but in case of a one-dimensional lattice, there
exists a relatively simple relation\cite{lgp88} between $\xi$ and $\langle
D(\omega) \rangle$, the disorder averaged density of states, 
\begin{eqnarray}\label{eqn thouless relation}
 \xi^{-1}(E) & = & \int\limits_{-\infty}^{\infty} \langle D(\varepsilon) \rangle \, \log\left|E-\varepsilon\right| \: d\varepsilon .
\end{eqnarray}
(Here and in the following, we choose units where $t=1$ and measure $\xi$ in units of the lattice spacing.)  \\
In the strong disorder limit, $\langle D(\varepsilon) \rangle$ can be replaced by $p_{\textrm{A}}(\varepsilon+\mu)$, so that the inverse localization length at the Fermi level $(E=0)$ is given by
\begin{eqnarray}\label{eqn simplified thouless}
 \xi^{-1} & = & \int\limits_{-\infty}^{\infty} p_{\textrm{A}}(\varepsilon) \, \log\left|\varepsilon-\mu\right| \, d\varepsilon \quad .
\end{eqnarray}
\begin{figure}
 \begin{center}
  \subfigure[]{\includegraphics[scale=0.4]{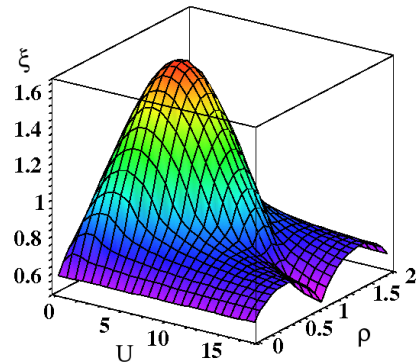} \label{fig loc length 3d atomic}}
  \subfigure[]{\includegraphics[scale=0.4]{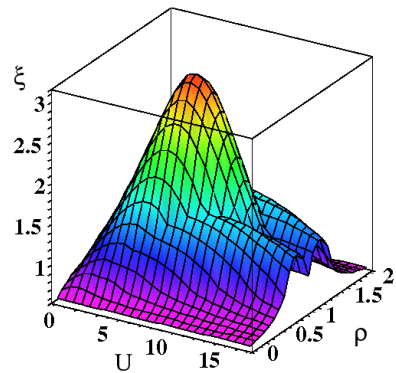} \label{fig loc length 3d Hartree}}
  \caption{(Color online) Localization length $\xi$ at the Fermi level as a function of repulsion $U$ and lattice filling $\rho$ for $\Delta=15$: (a) atomic-limit approximation and (b) Hartree-Fock approximation.}\label{fig loc length 3d}
 \end{center}
\end{figure}
A plot of $\xi$ as a function of $U$ and $\rho$, for fixed disorder strength $\Delta=15$, is shown in Fig. \ref{fig loc length 3d atomic}. It can be seen that for each given filling factor, the localization length exhibits a pronounced maximum. This maximum can be calculated to appear at
\begin{eqnarray}
 U_{\xi}^{\textrm{A}} & = & \frac{\Delta}{3}\left(\sqrt{1 + 3\rho (2-\rho)} - 1\right) \nonumber \\
 & \approx & \frac{\Delta}{2} \rho (2-\rho) \, + \, \mathcal{O}(\rho^{2} (2-\rho)^{2}) \label{eqn max U loc atomic limit}\: .
\end{eqnarray}
The reason for this nonmonotonic behavior is simple: A weak to
intermediate repulsion $U$ changes the on-site energy distribution from a rectangular distribution to a narrower one by shifting low on-site energies toward the Fermi level (screening), see
Fig. \ref{fig prob distribution atomic}. In contrast, a very strong on-site 
repulsion enhances the localization by a large broadening of the probability density. Therefore, in between, there will be
some value $U_{\xi}^{\textrm{A}}$ for which the screening is optimal and the 
localization length acquires a maximum. Such behavior is expected, since
a strong repulsion effectively suppresses hopping processes and leads to an accumulation of spectral weight in the upper Hubbard band. \\
In Fig. \ref{fig loc compare}, the localization length $\xi$ is shown as a
function of $U$ for $\rho=\frac{1}{2}$ (quarter filling). A similar,
nonmonotonic behavior was also found in recent quantum Monte Carlo simulations\cite{cds07,tdhs05,ds03,dst99} and a most recent statistical dynamical mean field theory evaluation\cite{swa07} of the problem, where the conductivity and the inverse participation ratio\cite{sbs03} of a finite system were calculated, respectively. Identifying a maximum of conductivity with a maximum of localization length, we find in all cases a good qualitative agreement with our results. Furthermore, we even find a reasonable quantitative agreement with our results for the points of maximal delocalization, $U_{\xi}^{\rm A}$, Eq.~(\ref{eqn max U loc atomic limit}), and $U_{\xi}^{\rm H}$, defined in Eq.~(\ref{eqn max U loc Hartree}) below. Thus, there is a strong correlation between the degree of screening and the conductivity: optimal screening corresponds to maximal conductivity. In addition, our analytical results
indicate that much of this physics of the nonmonotonic behavior can already be understood on the level of a static screening approximation.\\
Our result appears to be at odds with the statement\cite{cds07} that screening alone cannot account for the nonmonotonic behavior of the conductivity. We will discuss this point in the next section. For that discussion, we need the variance of the renormalized on-site energies which is given by
\begin{eqnarray}
& & \!\!\! \sigma^{2}_{\textrm{A}} \: \equiv \: \langle \varepsilon^{2} \rangle_{\textrm{A}} - \langle \varepsilon \rangle^{2}_{\textrm{A}} \\
& & \!\!\! = \frac{\Delta^{2}}{12}\left\{ \begin{array}{rcl} 1 - \hat{\rho}\hat{U} + \hat{\rho}\hat{U}^{2} + 3\hat{U}^{3} & \textrm{,} & \rho < 1 \, , \, \hat{U} < \rho \\ 1 - 2\rho'\hat{U} + \hat{\rho}\hat{U}^{2} & \textrm{,} & \rho < 1 \, , \, \hat{U} \geq \rho \\ 1 - \hat{\rho}\hat{U} + \hat{\rho}\hat{U}^{2} + 3\hat{U}^{3} & \textrm{,} & \rho \geq 1 \, , \, \hat{U} < 2-\rho \\ 1 - 2\tilde{\rho}\hat{U} + \hat{\rho}\hat{U}^{2} & \textrm{,} & \rho \geq 1 \, , \, \hat{U} \geq 2-\rho, \end{array} \right. \label{eqn var atomic limit} \nonumber 
\end{eqnarray}
with
\begin{eqnarray}\label{eqn definition hat_rho}
 & \hat{U} \: = \: U/\Delta\: , \quad \hat{\rho} \: = \: 3\rho(2-\rho)\: , \quad \rho' \: = \: 3\rho(1-\rho)\:, & \nonumber \\
 & \tilde{\rho} \: = \: 3\left(\frac{1}{4}-(\rho-\frac{3}{2})^{2}\right) . &
\end{eqnarray}
Here, $\sigma^{2}_{\textrm{A}}$ has a minimum at
\begin{eqnarray}\label{eqn min U var atomic}
U_{\sigma}^{\textrm{A}} = \left\{ \begin{array}{ccl} \frac{\Delta}{9}\left(\sqrt{\hat{\rho}^{2} + 9\hat{\rho}} - \hat{\rho}\right) &\, , \, & \frac{1}{3} \lesssim \rho \lesssim \frac{5}{3} \\ \Delta\frac{|1-\rho|}{1+|1-\rho|} &\, , \, & \textrm{otherwise}, \end{array} \right.  
\end{eqnarray}
which can be seen in Fig. \ref{fig var compare} for $\rho=\frac{1}{2}$ and $\Delta=15$. \\~\\
So far, our calculations were restricted to one dimension because a
generalization of relation (\ref{eqn thouless relation}) to two or three
dimensions is, in general, not possible. However, in the limit of strong
disorder, $G_{ij}(E)$ can be calculated in good approximation by taking into
account only the direct path\cite{mk92} from $i$ to $j$. Therefore, 
Eq.~(\ref{eqn def loc length}) can be considered as a one-dimensional problem, 
leading again to Eqs.~(\ref{eqn thouless relation}) and 
(\ref{eqn simplified thouless}), respectively. Hence, we conjecture 
that the general result, i.e., the nonmonotonic behavior of localization, holds also in two and three dimensions.

\section{Site-dependent Hartree-Fock approximation}
\begin{figure}
 \begin{center}
  \subfigure[]{\includegraphics[scale=0.2]{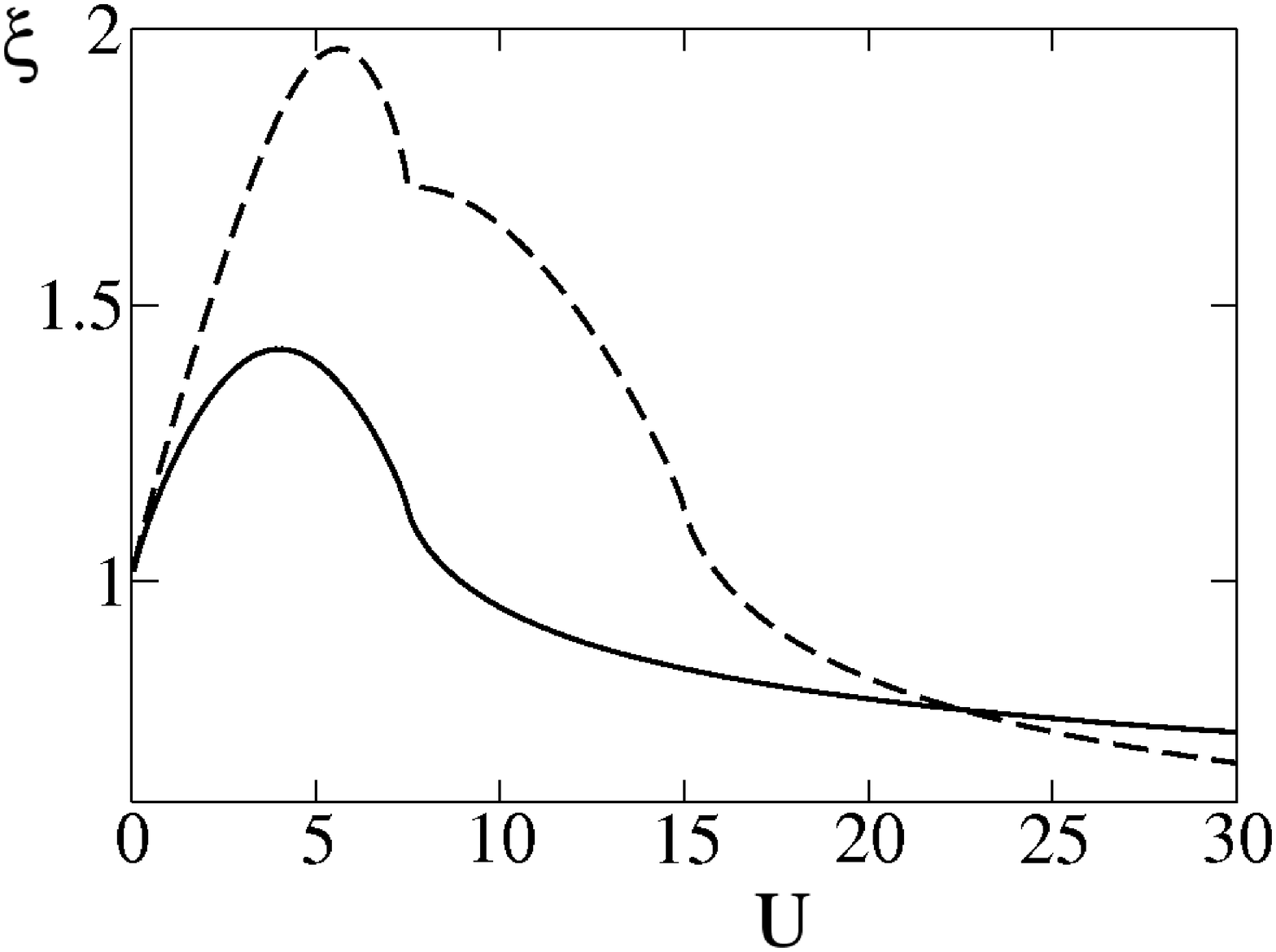}\label{fig loc compare}} \hfil
  \subfigure[]{\includegraphics[scale=0.2]{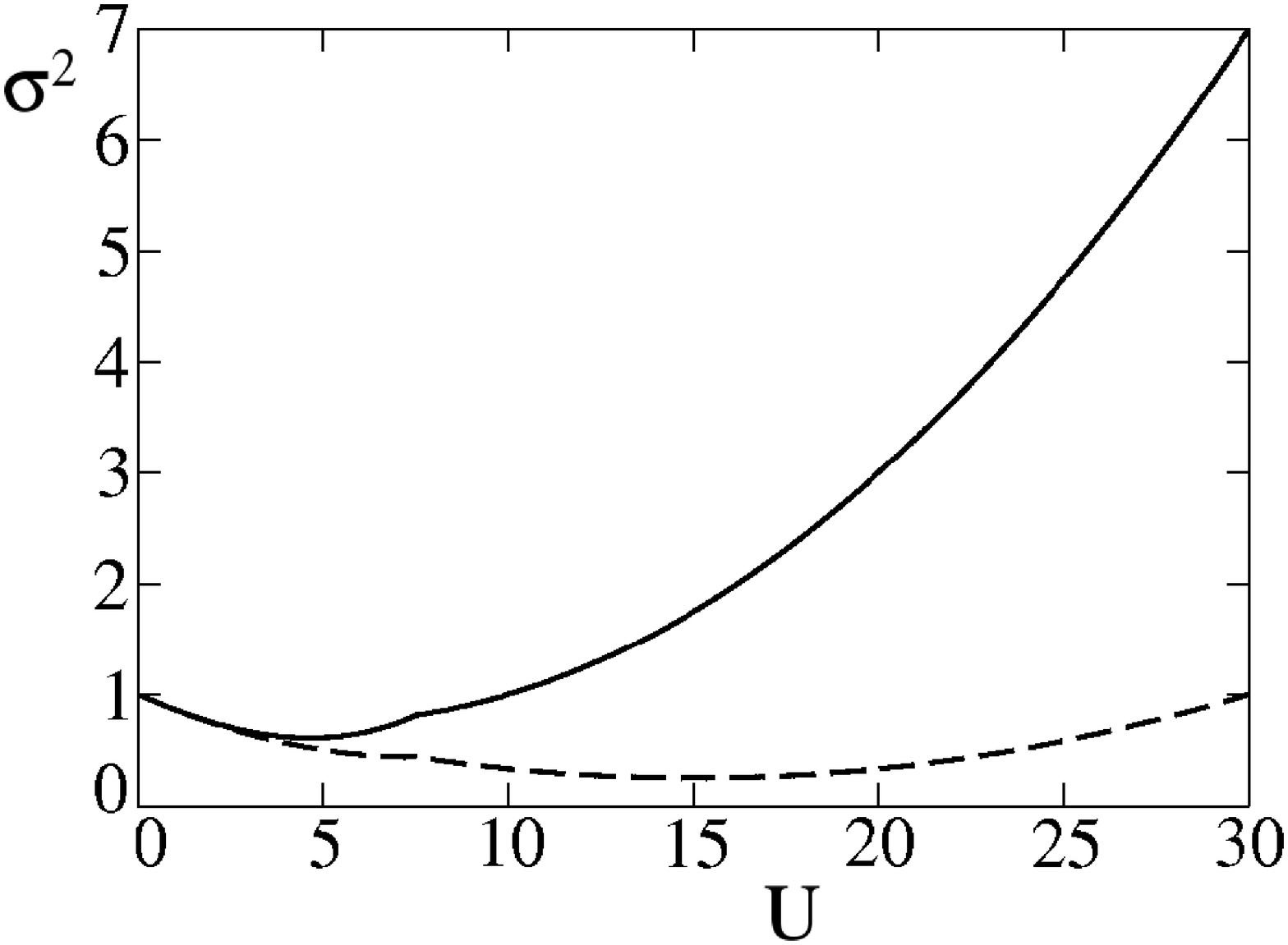}\label{fig var compare}}
  \caption{(a) Localization length $\xi$ at the Fermi level, normalized by its noninteracting value, as a function of the on-site repulsion $U$ for $\Delta=15$ and $\rho=\frac{1}{2}$. (b) Variance of the renormalized on-site energy distribution, also normalized by its noninteracting value. The solid curves show the results for the atomic-limit approximation and the dashed curves are the Hartree-Fock results.}\label{fig compare}
 \end{center}
\end{figure}
The second model which will be discussed in this paper is the site-dependent
Hartree-Fock approximation.\cite{cds07,her01} In this single-particle approximation, 
each site has a single renormalized energy level, given by 
\begin{eqnarray}\label{eqn renormalized energies}
\varepsilon_{i}^{\phantom{\dagger}} & \mapsto & \varepsilon_{i}^{\phantom{\dagger}} + \frac{U}{2} \langle n_{i}^{\phantom{\dagger}}\rangle_{0},
\end{eqnarray}
which in Eq.~(\ref{original Hamiltonian}) corresponds to the replacement,
\begin{eqnarray}
& & \!\!\! U n_{i\uparrow}^{\phantom{\dagger}} n_{i\downarrow}^{\phantom{\dagger}} \: \rightarrow \: U \left( \langle n_{i\uparrow}^{\phantom{\dagger}}\rangle\, n_{i\downarrow}^{\phantom{\dagger}} + \langle n_{i\downarrow}^{\phantom{\dagger}}\rangle\, n_{i\uparrow}^{\phantom{\dagger}} \right) \nonumber \\ 
 & &\!\!\! \quad  \rightarrow\:  \frac{U}{2} \langle n_{i}^{\phantom{\dagger}}\rangle \left(n_{i\downarrow}^{\phantom{\dagger}} +  n_{i\uparrow}^{\phantom{\dagger}} \right) \: \approx \: \frac{U}{2} \langle n_{i}^{\phantom{\dagger}}\rangle_{0} \left(n_{i\downarrow}^{\phantom{\dagger}} +  n_{i\uparrow}^{\phantom{\dagger}} \right) . \qquad  
\end{eqnarray}
Here, the absence of any kind of magnetization was assumed. Note that for
a local, energy independent interaction $U$, the Hartree and the 
Hartree-Fock approximations are identical.
The average on-site occupation $\langle n_{i}^{\phantom{\dagger}}\rangle$ was
taken to be the one of the atomic ground state, $\langle
n_{i}^{\phantom{\dagger}}\rangle_{0}$, Eq.~(\ref{eqn occupation numbers atomic
  limit}), according to the assumption of a stable atomic configuration.  
\\
As in the atomic-limit approximation, the shift of the occupied on-site
energies leads to a renormalized distribution, with probability function
$p_{\textrm{H}}(\varepsilon)$ (Fig. \ref{fig prob distribution hartree}). The
screening effect now is even more pronounced because all singly occupied
states are shifted by $\frac{U}{2}$ yielding a smaller width and a stronger
increase of the probability to find a state around the Fermi level. The
resulting effective single-particle Hamiltonian is again given by
Eq.~(\ref{atomic Hamiltonian}), however, with the probability function $p_{\textrm{H}}(\varepsilon)$. \\
The corresponding plots for the localization length in Hartree-Fock 
approximation are shown in Figs. \ref{fig loc length 3d Hartree} and \ref{fig loc compare}, respectively. Again, a pronounced maximum of $\xi$ arises. In this case, it is found at the value
\begin{eqnarray}\label{eqn max U loc Hartree} 
  U_{\xi}^{\textrm{H}} & = & \frac{\Delta}{2}\rho (2-\rho) ,
\end{eqnarray}
which up to order $\mathcal{O}\left(\rho^2\cdot(2-\rho)^2\right)$ coincides with the result from the atomic-limit approximation Eq.~(\ref{eqn max U loc atomic limit}). The increase of the localization length is considerably more pronounced in the Hartree-Fock approximation due to its narrower probability distribution and especially its larger probability density around the Fermi level. In both approaches, the effect of screening becomes stronger with decreasing disorder. However, for small values of $\Delta$, the stability of the atomic ground state becomes doubtful. \\ 
The average and the variance of the renormalized on-site energy distribution
are, respectively,
\begin{eqnarray}
 \langle \varepsilon \rangle_{\textrm{H}} \!\! & = & \langle \varepsilon \rangle_{\textrm{A}} \: = \: \frac{1}{2}\,\rho\, U \, , \\
 \sigma^{2}_{\textrm{H}} & = & \frac{\Delta^{2}}{12}\left\{ \begin{array}{rcl}
 1 - \hat{\rho}\hat{U} + \hat{\rho}\hat{U}^{2} & \!\! ,\! & \rho < 1 \, , \, \hat{U} < \rho \\ 1 - 2\rho'\hat{U} + \rho'\hat{U}^{2} & \!\! ,\! & \rho < 1 \, , \, \hat{U} \geq \rho \\ 1 - \hat{\rho}\hat{U} + \hat{\rho}\hat{U}^{2} & \!\! ,\! & \rho \geq 1 \, , \, \hat{U} < 2-\rho \\ 1 - 2\tilde{\rho}\hat{U} + \tilde{\rho}\hat{U}^{2} & \!\! ,\! & \rho \geq 1 \, , \, \hat{U} \geq 2-\rho, \end{array} \right. \qquad \!\! \label{eqn var Hartree}
\end{eqnarray}
with the same abbreviations as in Eq.~(\ref{eqn definition hat_rho}). 
Its minimum is found at the value
\begin{eqnarray}\label{eqn min U var Hartree}
U_{\sigma}^{\textrm{H}} = \left\{ \begin{array}{rcl} \Delta/2 &\, , \, & \frac{2}{3}<\rho<\frac{4}{3} \\ \Delta &\, , \, & \textrm{else}. \end{array} \right. 
\end{eqnarray}
\begin{figure}
 \begin{center}
  \subfigure[]{\includegraphics[scale=0.2]{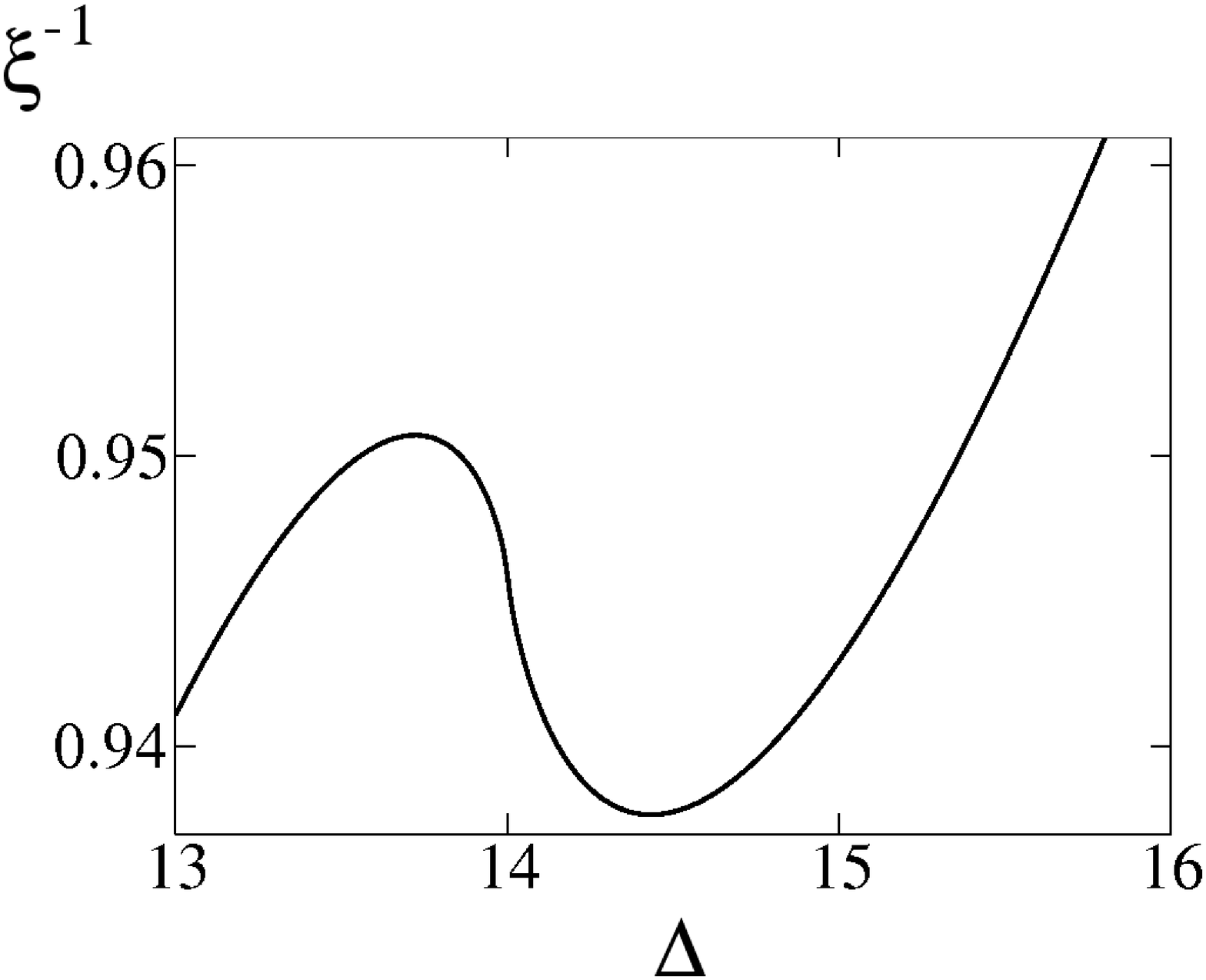}} \hfil
  \subfigure[]{\includegraphics[scale=0.2]{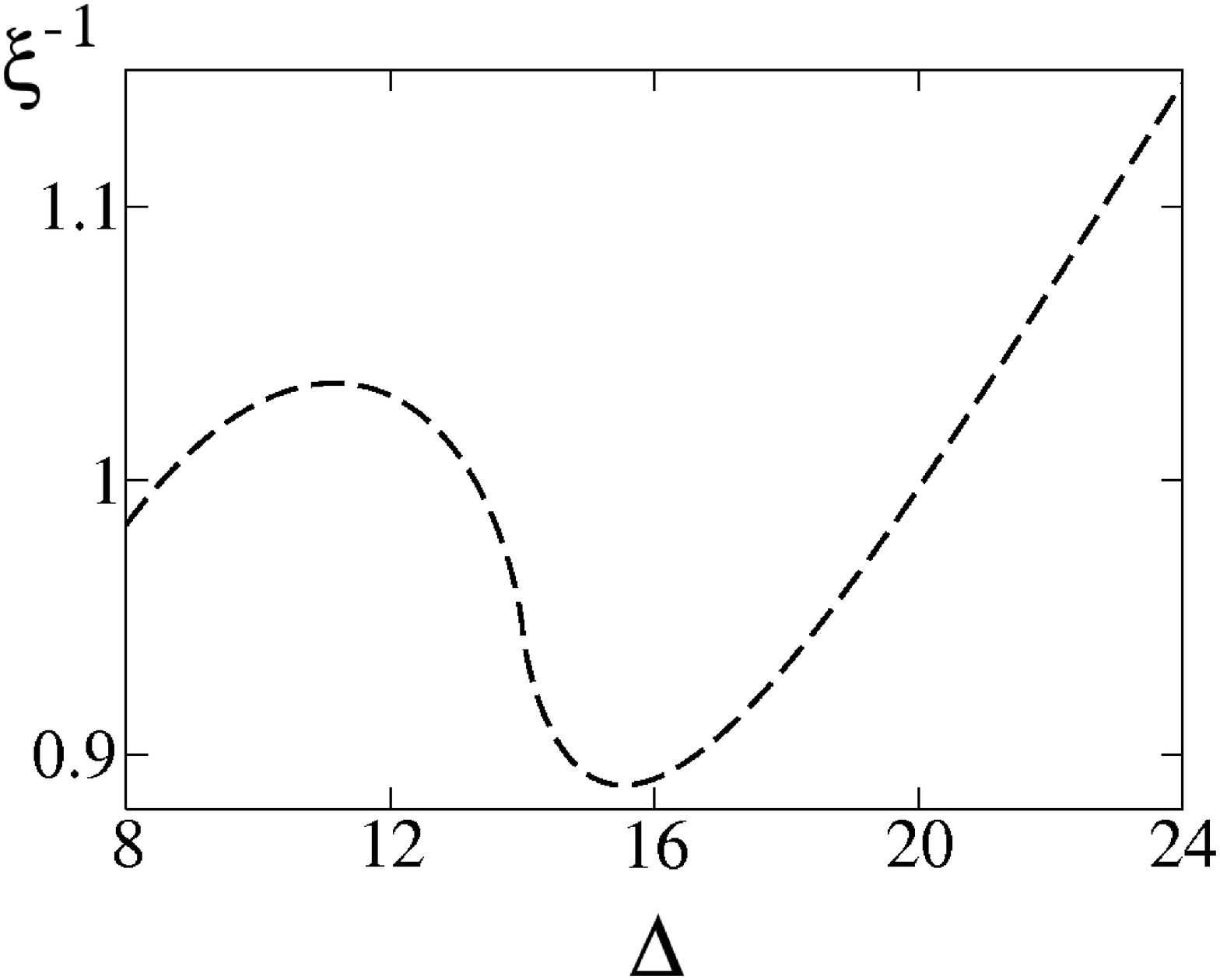}}
  \caption{Inverse localization length $\xi^{-1}$ as a function of $\Delta$ for $\rho=\frac{1}{2}$ and (a) $U=7$ in the atomic-limit approximation and (b) $U=14$ in the Hartree-Fock approximation, respectively.}\label{fig ipr}
 \end{center}
\end{figure} 
\\ As mentioned above, it was argued in Ref.~\onlinecite{cds07} that the picture of screening would be too primitive to explain the nonmonotonic behavior and the evidence for a metallic state found in the conductivity simulations by varying the repulsion strength $U$. The argumentation was based on the observation, which the variance was a featureless, monotonically decreasing function of $U$ around the transition point. Our results show that this
reasoning is generally not conclusive. Although the static, single-particle treatment does not allow for the occurrence of a metallic state, we find a strong enhancement and nonmonotonic behavior of the localization length $\xi$ as function of $U$, whereas the variance is also only a monotonically decreasing function around the point of maximal delocalization.
For strong disorder and a distribution which is not characterized by a single parameter (like in the present case, cf. Fig.~\ref{fig prob distribution}), there is no simple relation between $\xi$ and
the variance $\sigma^{2}$. Especially for lower fillings, our results, Eqs.~(\ref{eqn max U loc atomic limit}), (\ref{eqn min U var atomic}) and (\ref{eqn max U loc Hartree}), (\ref{eqn min U var Hartree}), respectively, show that
the values of $U$ for which the maximum of $\xi$ and the minimum of $\sigma^{2}$ occur, can be separated systematically. Moreover, we find that the atomic-limit approximation and the Hartree-Fock approximation do yield close values of $\xi$, although the variances can differ strongly (see Fig. \ref{fig compare}). \\
In Refs.~\onlinecite{ht04} and ~\onlinecite{tdhs05}, the inverse participation ratio was calculated as a function of the disorder strength $\Delta$, and a nonmonotonic behavior with evidence for a metallic state was found. It was argued there that the screening picture would necessarily predict a monotonic increase of the inverse participation ratio with increasing $\Delta$, excluding screening as a possible explanation. Our model, contrary to that statement, exhibits such nonmonotonic behavior as well, as shown in Fig. \ref{fig ipr}, where $\xi^{-1}$ is plotted as a function of the disorder strength $\Delta$, for some fixed values of $\rho$ and $U$. This nonmonotonicity is caused by the crossover from the regime of disorder screening by interaction, $\Delta \gg U$, to the regime of interaction-reduced hopping, $\Delta \ll U$, which is controlled by the parameter $U/\Delta$. However, the exact position of the nonmonotonicity depends also on $t/\Delta$, $\rho$, and the probability function.

\section{Conclusion}
We examined the effect of static disorder screening by on-site repulsion in the
one-dimensional Hubbard-Anderson model for strong disorder. We presented two
different approximation schemes by absorbing the interactions into a
redefinition of the single-particle on-site energies. In both approaches a renormalized probability distribution with an enhanced probability of
finding site energies close to the Fermi level was obtained. We calculated the
localization length at the Fermi energy for these single-particle problems
and found a pronounced maximum of the localization length for some
intermediate value of the repulsion strength. This can be understood as a consequence of the fact that the increase of the localization length $\xi$ for 
small $U$ ($U<U_{\xi}$) and the decrease of $\xi$ for large $U$
($U>U_{\xi}$) have different physical origins, namely disorder screening 
and reduced hopping, respectively.  Similarly, a change of the
''bare'' disorder $\Delta$, for fixed repulsion, resulted in a nontrivial,
nonmonotic dependence of the localization length on the disorder strength. In
contrast to the case of weak disorder, we found no significant correlation
between the variance of the effective on-site energy distribution and the
localization length. Our results, especially in the case of the Hartree-Fock approximation, are in
qualitative and to some degree even quantitative agreement with recent
numerical studies. By our analytic approach, it was possible to investigate
the static screening effects separately from dynamical (inelastic)
processes. \\
We gave an argument that the same behavior should also be found for
strong disorder in two and three dimensions.\cite{fussnote5} In three dimensions there might 
be an interesting possibility of an interaction induced metal-insulator
transition. Such a possibility is based on the assumption that our results,
obtained in the strongly localized regime, can be extrapolated up to the
mobility edge. Under this assumption, a noninteracting Anderson-localized
system, whose Fermi energy is sufficiently close to the mobility edge, will
become metallic upon switching on interactions by shifting the mobility edge
across the Fermi level. Furthermore, when the interactions exceed a certain
strength, the system would reenter the insulating phase in analogy to the
nonmonotonic behavior of the localization length in one
dimension.\cite{fussnote2} (Let us emphasize that throughout this
paper we do not discuss the case of half filling, with its characteristic
Mott's physics.\cite{adak07, fussnote3}) A similar, and experimentally more
relevant, effect could also happen under a change of the bare disorder
$\Delta$, as suggested by Fig. \ref{fig ipr}, since the relevant dimensionless
parameter is $U/\Delta$. 

\acknowledgments
This work was supported in part by the Deutsche Forschungsgemeinschaft
through SFB 608. P.H. acknowledges additional support by Deutscher
Akademischer Austausch Dienst (DAAD). B.S. acknowledges the
hospitality of Bonn University where the present work was
initiated.


\begin{thebibliography}{99}
\bibitem{fussnote4} At half filling and strong repulsion, the chemical potential lies in the gap of the two Hubbard bands. To avoid an unnecessary case differentiation, in that case, we defined $\mu=U-\Delta/2$. $\mu$ defined in this way is at the top of the gap, in contrast to the more common definition, $\mu=U/2$, at the center of the gap. Physical quantities are of course not affected by this definition.
\end{thebibliography}
\end{document}